\documentclass[12pt,preprint]{aastex}
\begin{document}

\parindent=1.0cm
\title {Red Supergiants in the Disk of M81: Tracing the Spatial Distribution of Star 
Formation 25 Million Years in the Past \altaffilmark{1}}

\author{T. J. Davidge \altaffilmark{2}}

\affil{Herzberg Institute of Astrophysics,
\\National Research Council of Canada, 5071 West Saanich Road,
\\Victoria, B.C. Canada V9E 2E7\\ {\it email: tim.davidge@nrc.ca}}

\altaffiltext{1}{This publication makes use of data products
from the Two Micron All Sky Survey, which is a joint project of the University of
Massachusetts and the Infrared Processing and Analysis Center/California
Institute of Technology, funded by the National Aeronautics and Space Administration
and the National Science Foundation.}

\altaffiltext{2}{Visiting Astronomer, Canada-France-Hawaii
Telescope, which is operated by the National Research Council of Canada,
the Centre National de la Recherche Scientifique, and the University of
Hawaii.}

\begin{abstract}

	Near-infrared images, obtained with the CFHTIR imager on the Canada-France-Hawaii 
Telescope, are used to investigate the brightest red stars in the disk of the 
nearby spiral galaxy M81. Red supergiants (RSGs) form a well-defined sequence on the 
color-magnitude diagrams (CMDs) that peaks near M$_K = -11.5$; 
RSGs with this peak brightness are seen throughout all fields that 
were studied, indicating that star formation occured over a large part of 
the M81 disk only $\sim 10$ Myr in the past. The number of RSGs per unit integrated 
$K-$band light is compared at various locations in the disk. The number 
density of bright RSGs is similar in three of four fields, indicating that the 
bright RSGs tend to be well mixed with the older stellar populations that dominate the 
integrated light in the $K-$band. However, the density of bright RSGs 
in a northern disk field is $\sim 2 \times$ higher than average, suggesting that 
the SFR in this part of the disk was higher than average $10 - 25$ Myr in the past. 
The northern disk field contains areas of on-going star formation, and it is 
suggested that it is a region of prolonged star-forming activity. 
The number density of RSGs that formed during the past $10 - 25$ Myr 
at galactocentric distances between $\sim 4$ and 7 kpc is also comparable with that 
which formed between $\sim 7$ and 10 kpc. We conclude that star-forming activity in 
M81 during the past $10 - 25$ Myr (1) was distributed over a larger fraction of 
the disk than it is at the present day, and (2) was not restricted to 
a given radial interval, but was distributed in a manner that closely 
followed the stellar mass profile. Star counts indicate that the mean SFR of M81 
between 10 and 25 Myr in the past was $\sim 0.1$ M$_{\odot}$ year$^{-1}$, which is not 
greatly different from the present day SFR estimated from H$\alpha$ and FUV emission. 

\end{abstract}


\section{INTRODUCTION}

	It is currently thought that large galaxies 
formed through the progressive merging of smaller units. The 
rate of galaxy assembly via merger activity was undoubtedly much higher during 
early epochs than at the present day, due to the higher density of objects. The 
intense merger activity that likely occured during early epochs is associated with 
the formation of spheroidal systems, and the luminosity-weighted old ages of 
the central regions of bulges (e.g. Sarzi et al. 2005) is consistent with this 
expectation. In contrast, disk formation may have been a more extended process, that 
continued to comparatively recent epochs (e.g. Hammer et al. 2004).

	If the hierarchal merging model is correct 
then we would expect to see evidence of significant interactions between 
nearby spiral galaxies and their satellites. 
In fact, there are indications that the Milky-Way has interacted with, or is interacting
with, some of its satellites. Tidal streams and related structures have been found in the 
outer regions of the Milky-Way (e.g. Ibata et al. 2001; Martinez-Delgado et al. 2001; 
Ibata et al. 2003), and it has been suggested that some Galactic globular clusters 
formed in dwarf galaxies (Bellazzini, Ferraro, \& Ibata 2003; MacKey \& 
Gilmore 2004), that have since been accreted by the Milky-Way. 
It has been suggested that some objects identified as globular clusters, but having 
extended sizes and/or evidence for complex star-forming histories, may instead be 
the remnants of the central cores of disrupted dwarf galaxies (Bekki \& Freeman 2003; 
Mackey \& van den Bergh 2005). Moving groups of metal-poor stars in the solar 
neighborhood may also be the dispersed remains of dwarf galaxies that were 
accreted along the disk plane (e.g. Meza et al. 2005).

	Studies of the closest large external spiral galaxy have also found 
evidence of interactions between M31 and its satellites. Tidal structures have 
been discovered in the outer regions of M31 (e.g. Ibata et al. 2001; Zucker et 
al. 2004), and there are indications that many of the globular clusters in M31 may have 
formed in a system that had a chemical enrichment history that was distinct from 
that of the main body of the present-day galaxy (Beasley et al. 2005). There is also 
a population of objects that may be the remnants of disrupted 
systems (e.g. Bekki \& Chiba 2004; Huxor et al. 2005). As has been suggested for the 
Milky-Way (e.g. Gilmore, Wyse, \& Norris 2002), the interactions that generated 
the tidal streams in M31 likely occured well in the past, as they 
do not contain stars younger than 2.5 Gyr (Ferguson et al. 2005). 

	To study more recent interactions it is necessary to 
move beyond the Local Group. The star-forming galaxies NGC 3077 and M82 
have been interacting with M81, which is the closest Sb galaxy after M31, for at least a 
few 100 Myr (e.g. Brouillet et al. 1991; Yun, Ho, \& Lo 1994) and possibly 
for longer than 1 Gyr (Parmentier, de Grijs, \& 
Gilmore 2003). The interactions continue to this day, with gas streams linking M81, 
NGC 3077, and M82 (Yun et al. 1994) and other members of the group (e.g. Boyce 
et al. 2001). Tidal interactions have affected the gas distribution in M81, with HI 
emerging from the south and east sides of the galaxy.

	The interactions that induced star formation in M82 and NGC 3077 would also 
be expected to fuel elevated star formation rates (SFRs) 
in M81. Gordon et al. (2004) reviewed multi-wavelength 
observations of M81, spanning the far-ultraviolet to the radio, to 
compare various star formation indicators. These indicators 
probe relatively recent (i.e. within the past $\sim 10$ Myr) star-forming activity, and 
indicate that star formation is presently concentrated at intermediate galactocentric 
radii in the northern, western, and southern portions of M81. The SFR at the present day 
in M81 is $\sim 0.3$ M$_{\odot}$ year$^{-1}$ (Bell \& Kennicutt 2001), which is 
roughly a third that in M31 (Williams 2003).

	Efforts to probe star-forming activity in M81 during earlier epochs are 
of interest, as the interactions with NGC 3077 and M82 were likely more intense 
in the past. For example, the age distribution of star clusters in M82 suggest 
that the SFR in that galaxy peaked $\sim 1$ Gyr in the past (Parmentier et al. 
2003). In fact, evidence for vigorous star formation in M81 during the past 1 Gyr 
is seen. Chandar, Ford, \& Tsvetanov (2001) find compact clusters with blue B-V colors that 
extend out to galactocentric distances of at least R$_{GC} \sim 10$kpc, and Chandar, 
Tsvetanov, \& Ford (2001) estimate that these clusters have ages between 6 and 600 Myr. 
Swartz et al. (2003) find a population of centrally-concentrated low-mass x-ray binaries 
with an age of $\sim 400$ Myr, indicating that there was a significant episode 
of star formation in the M81 bulge during the past few hundred million years; 
residual star-forming activity might still be occuring near 
the center of M81 (Davidge \& Courteau 1999).

	Interactions between a large galaxy and its companions might also 
result in the formation of dwarf galaxies if gas that is pulled from the disk 
by tidal interactions cools sufficiently to form stars 
(e.g. Kroupa 1998), and it has been suggested that 
some dwarf galaxies that are close to M81 may have 
formed in this way. Based on its dynamical age, van Driel et al. (1998) 
argue that the Garland formed out of tidal material, while Boyce et al. (2001) suggest 
that some dwarf systems in the M81 group that are embedded in HI streams may 
have recently condensed out of this material. Makarova et al. (2002) 
examine the resolved stellar contents of four 
systems associated with tidal streams in the M81 group, and conclude that 
these may have condensed from tidal material during the past few hundred Myr. 
Sun et al. (2005) find possible evidence for a diffuse stellar component in the 
M81 group, which is likely tidal in origin.

	As noted by Chandar et al. (2002), that the youngest 
compact clusters in the disk of M81 have ages of only 6 
Myr suggests that vigorous star formation occured in M81 during the not-to-distant past. 
The star-forming history of a galaxy during the past few tens of millions of years 
can be probed by investigating the properties of RSGs. 
In the current paper, near-infrared images obtained with two instruments 
on the Canada-France-Hawaii Telescope (CFHT) are used to investigate 
the RSG and bright AGB content in the disk of M81. 
Observations in the near-infrared are important for studying and characterizing 
RSGs and AGB stars, as these objects form near-vertical 
sequences on infrared CMDs. Not only does this increase the contrast 
with respect to the fainter, predominantly blue, stellar substrate 
of the M81 disk, but it also simplifies the task of 
identifying the peak brightness of these components.

	The main dataset used in this investigation was obtained with the CFHTIR imager, 
and consists of moderately deep images of four different locations in the M81 disk at 
intermediate galactocentric radii. Supplemental observations of a sub-region within 
one of the CFHTIR fields were obtained with the Canada-France-Hawaii Telescope (CFHT) 
adaptive optics (AO) system and KIR imager. The observations and the procedures used to 
process the data are described in \S 2. The photometric measurements and the CMDs 
are discussed in \S 3, while the specific frequency of RSGs and AGB 
stars, which is the number of these stars per unit integrated 
brightness per magnitude interval, is examined in \S 4. A summary and discussion of the 
results follows in \S 5.

\section{OBSERVATIONS \& REDUCTIONS}

\subsection{CFHTIR Data}

	Four fields that sample the outer disk of M81 were observed with the CFHTIR imager 
during the nights of November 21, 22, and 23 2002. The CFHTIR contains a $1024 \times 1024$ 
HgCdTe array, with each pixel subtending 0.22 arcsec on a side. 
A $3.8 \times 3.8$ arcmin$^2$ field is thus imaged during each exposure, and this 
corresponds to a projected area of $3.9 \times 3.9$ kpc$^2$ at the distance of M81, 
which is assumed to be 3.55 Mpc (Freedman et al. 2001) throughout this study.

	The CFHTIR images sample the spiral arms of M81 at intermediate galactocentric 
distances. The central co-ordinates of the fields are listed in Table 1, while 
their locations are indicated in Figure 1, which shows a portion of the $K-$band image 
of M81 from the 2MASS Large Galaxy Atlas (Jarrett et al. 2003). The data for each field 
were recorded as a series of $30 - 60$ second exposures at each point in a $2 \times 2$ 
square dither pattern. All fields were observed through $J, H,$ and $K'$ filters, with 
a total exposure time of 12 minutes filter$^{-1}$. The West field was also 
observed through $K$ continuum and CO filters, with a total exposure time of 12 minutes 
for the first filter and 18 minutes for the second. 

	The basic steps in the processing sequence were (1) dark 
subtraction, (2) division by flat-field frames, and (3) the removal of interference 
fringes and thermal emission from warm objects along the light path. The processed 
images were then spatially registered and median-combined on a field-by-field and 
filter-by-filter basis, after adjusting for variations in the DC sky level. 
The combined images of each field were then trimmed to the area common to all exposures of 
that field to create the final images used in the photometric analysis.
Stars in the final images have FWHM between 0.7 and 1.1 arcsec FWHM.

\subsection{AO Data}

	The area around the star GSC 04383--00308, which is at the 
center of the CFHTIR M81 West field, was observed with the CFHT AO system and KIR imager 
during the night of March 10 2001. The KIR imager contains a $1024 \times 1024$ 
HgCdTe array, and each pixel subtends 0.034 arcsec on a side. 
The $34.4 \times 34.4$ arcsec$^2$ field that is imaged in each exposure 
covers a projected area of $0.64 \times 0.64$ kpc$^2$ at the distance of M81.

	Images were recorded through $J$, $H$, and $K'$ filters, with a total exposure 
time of 20 minutes filter$^{-1}$. The CFHT AO system uses natural guide stars, 
and GSC 04383--00308 was used as the reference source for image 
corrections. The AO data were reduced using the procedures 
described in \S 2.1. The seeing was mediocre when these data were recorded, and 
was estimated to be 0.9 -- 1.0 arcsec in $V$. As a result, the angular resolution 
of the final images is poor by AO standards, with FWHM $\sim 0.3$ arcsec in each filter. 
Still, this is far superior to the angular resolution of the CFHTIR data, and the 
AO-corrected images are used to check if the star counts at the faint end of the 
CFHTIR data are affected by crowding. 

\section{RESULTS}

	The photometric measurements were obtained 
with the PSF-fitting routine ALLSTAR (Stetson \& Harris 1988), using PSFs and star 
lists obtained from tasks in the DAOPHOT (Stetson 1987) package. The photometric 
calibration was defined using stars in the 2MASS Point Source 
Catalogue (Cutri et al. 2003). Only stars with $K < 15$ were used as calibrators, 
as this is the magnitude range where the errors in the photometry are less than 
10\% in $K$. The field-to-field scatter in the 
zeropoints computed individually for the North, South, and West fields is modest, 
with standard deviations about the mean zeropoint of $\pm 0.002$ mag in $J$, $\pm 
0.019$ mag in $H$, and $\pm 0.042$ mag in $K$. The zeropoints for the 
East field are $\sim 0.15$ magnitudes brighter than for the other fields, suggesting 
that thin clouds may have been present when these data were recorded.

	The $(K, H-K)$ and $(K, J-K)$ CMDs of the CFHTIR fields are shown in Figures 2 
and 3. The CFHTIR CMDs are similar in appearance to those of other star-forming galaxies 
(e.g. Davidge 2005; 2006), and four groups of sources are present: (1) 
foreground stars, (2) star clusters, some of which have been 
identified previously (\S 3.2), (3) RSGs, and (4) the brightest 
AGB stars. Foreground stars dominate the CMDs when $K < 16$, while 
globular clusters may be present over a range of brightnesses (see below). 
RSGs populate the vertical red sequence that runs from $K = 17$ to $K = 19$, and the 
brightest AGB stars dominate the red clump of objects with $K > 19$. 
The CMDs of the various fields have similar appearances, and this foreshadows the 
main conclusion of this study, which is that the brightest evolved red stars tend 
to be uniformly mixed throughout the disk of M81 (\S 4).

	Artificial star experiments were run to 
assess sample completeness in the CFHTIR data, which can affect 
field-to-field comparisons of objects near the faint limit. 
Artificial stars were assigned colors and brightnesses that are similar to the main 
locus of objects in the CMDs. These experiments reveal field-to-field differences 
in the magnitude at which incompleteness sets in, which are due to variations in 
image quality and sky transparency, with the latter likely being the dominant issue in the 
East field  (see above). If an artificial star is assumed to be recovered if it is 
detected in both $H$ and $K$, which is the criterion used to construct 
the LFs that serve as the basis for comparing star counts in \S 4, then 
50 \% completeness occurs when $K \sim 19.5$ in the North and West 
fields. For comparison, 50\% completeness occurs at $K \sim 19$ in the South 
field, and $K \sim 18.5$ in the East field.

	The $(K, H-K)$ and $(K, J-K)$ CMDs of the AO field near the 
center of the CFHTIR West field are shown in the right hand panels of Figures 2 and 3. 
The AO data sample a much smaller field of view than the CFHTIR data, and so the number 
of sources in the AO-based CMDs is smaller than in the CFHTIR data. One impact of the 
modest sky coverage is that there is an absence of the brightest -- and rarest -- RSGs in 
the AO data, and the majority of objects in the right hand panels of Figures 2 and 3 
are AGB stars. 

	The difference in the brightnesses of objects that are common to the 
CFHTIR West the AO fields were computed, and the standard deviation about the mean was 
found to be $\pm 0.2$ magnitudes in all three filters. This is an upper limit to the 
scatter that is due to random photometric errors, as many of the brightest 
AGB stars are long period variables, with amplitudes of up to $\sim 1$ magnitude 
in the near-infrared, and the two datasets were recorded almost twenty months 
apart. In fact, the artificial star experiments predict that the random uncertainities 
in the CFHTIR data are $\pm 0.11$ magnitudes for stars with $K$ between 18 and 18.5, 
and the predicted dispersion in the AO data is comparable to this. Consequently,
much of the observed dispersion in the mean magnitude difference can be accounted for by 
random errors in the photometry. The comparatively good agreement between the measurements 
in the AO$+$KIR and CFHTIR datasets indicates that the photometry of AGB stars in the 
CFHTIR data is not affected significantly by crowding.

\subsection{Foreground Stars in the CFHTIR Images}

	The brightest RSGs in nearby galaxies have M$_K \sim -12$ 
(Rozanski \& Rowan-Robinson 1994), which corresponds to $K \sim 15.9$ 
at the adopted distance of M81. Few globular clusters are also brighter than M$_K \sim -12$ 
(e.g. Barmby, Huchra, \& Brodie 2001). Therefore, the majority of objects with $K < 16$ in 
the M81 CMDs are likely foreground stars, and this can be checked with star counts. 
Foreground stars will be uniformly distributed over degree scales on the sky, 
and their density can be measured at relatively large 
(i.e. a few tens of arcmin) angular offsets from M81. 
The density of objects with $K < 15$ more than 20 arcmin from the center of M81 in the 
2MASS Point Source Catalogue (Cutri et al. 2003) is $0.36 \pm 0.03$ 
arcmin$^{-2}$, where the uncertainty is the standard deviation of measurements made 
in different locations. This corresponds to 4 -- 5 stars per CFHTIR field. For comparison, 
there are 4 objects with $K < 15$ in the East, South, and West fields, as expected if 
these are foreground stars, although there is only 1 star with $K < 15$ in the North field.

	The CO measurements of the M81 West field have a faint limit of 
$K \sim 16.5$, and so can be used to provide supplementary 
insight into the nature of the brightest objects in that field. 
Late-type dwarfs, which are expected to dominate the foreground star populations, 
have CO between 0 and 0.1 (e.g. Frogel et al. 1975). The mean CO 
index for objects with $K < 16$ is $0.09 \pm 0.02$, where the quoted error is the 
uncertainty in the mean. Thus, the brightest objects have CO 
indices that are consistent with them being late-type dwarfs.

\subsection{Globular Clusters in the CFHTIR Images}

	The CFHTIR fields cover projected areas of $\sim 15$ kpc$^{2}$. Given this 
spatial coverage, it is perhaps not surprising that a modest number of globular clusters 
are present in these data. Globular clusters may masquerade as bright, evolved 
stars, and so we examine the number of clusters that might be present and their 
photometric properties. This is done using the sample of objects
identified by Perelmuter \& Racine (1995), who used a combination of morphology 
and spectral-energy distributions at visible wavelengths to identify candidate clusters.

	To identify candidate clusters it is assumed that the photometric properties of 
globular clusters in M81 are the same as those in M31, which appears to be the case 
based on existing spectroscopic (e.g. Perelmuter, Brodie, \& Huchra 1995; Schroder et al. 
2002), and photometric (Ma et al. 2005) information. 
The majority of globular clusters in the Milky-Way and 
M31 have $V-K$ between 2 and 2.5 (e.g. Barmby et al. 2000), 
and so the $V = 21$ faint limit of the Perelmuter \& Racine (1995) survey corresponds to 
$K \sim 18.5 - 19.0$, which is comparable to the faint limit of the CFHTIR data. 
Based on clusters in M31, the mean $J-K$ color of clusters in M81 will also be expected to 
be 0.67, with a standard deviation $\pm 0.13$ magnitudes (Barmby et al. 2000). 
Finally, if the M81 globular cluster luminosity function (GCLF) is like that in M31 
(e.g. Barmby et al. 2001), then the peak of the M81 GCLF should occur near $K = 
17.5$, with the majority of globular clusters having $K$ between 16 and 18.5. 
The vast majority of clusters in M81 will then likely (1) have brightnesses that 
fall within the detection limits of the CFHTIR data and (2) have $J-K < 1$, which is 
bluer than the brightest RSGs and AGB stars.

	The number of objects with $J-K$ falling within the $1\sigma$ limits of the 
M31 cluster color distribution, as computed by Barmby et al. (2000), was 
counted in each of the CFHTIR fields. A total cluster number was then found by 
multiplying the result by 1.3 to account for objects with colors outside of the $\pm 
1\sigma$ limits. Because the globular clusters in these fields are projected against -- 
or seen through -- the disk of M81 then they may have a range of reddenings. 
Two plausible extremes for $E(B-V)$ are considered: (1) $E(B-V) = 0.03$, 
which is the foreground value, and (2) $E(B-V) = 0.3$, which is the mean value 
for objects in the M81 disk computed by Perelmuter \& Racine (1995). With the lower 
reddening there is an average of $9 \pm 5$ objects per CFHTIR field, where the quoted 
uncertainty is the standard deviation about the mean. For the higher reddening the 
average number of objects is $23 \pm 12$. The second number is higher 
than the first because there is contamination from RSGs 
and AGB stars. For comparison, there are 4 Perelmuter \& Racine (1995) cluster 
candidates in the North and West fields, and 7 -- 8 candidates in the East and 
South fields. The number of blue objects found here are thus in rough agreement 
with those expected if only foreground reddening is assumed. The exact number of 
clusters notwithstanding, it appears that globular clusters can 
account for many of the blue sources in the CMDs in Figures 2 and 3, and so must be 
considered when studying the properties of the brightest evolved stars in these data. 

	Samples of spectroscopically confirmed globular clusters have been discussed by 
Brodie \& Huchra (1991), Perelmuter et al. (1995), and Schroder et al. (2002). Four of the 
globular clusters that were observed by Perelmuter et al. (1995) are in the East and 
West fields, and the infrared brightnesses and colors of these objects, along with 
selected entries from Table 4 of Perelmuter et al. (1995), are listed in Table 2. The 
colors and brightnesses in Table 2 have not been corrected for reddening.

	Despite having a range of spectroscopic metallicities, the entries in Table 2 
show only a modest spread in $J-K$ and $V-K$ colors, with values typical of 
those seen among globular clusters in other galaxies. While [Fe/H] for Id50415 and 
Id50552 differ by almost 3 dex, these objects have similar $V-K$ colors. 
The cluster Id50785 is bright enough to be detected in the CO 
observations, and the CO index is 0.19, with an estimated 
random error of $\pm 0.1$ magnitude. This is broadly consistent with the 
metallicity estimated for this cluster from visible light spectra.

\subsection{Red Supergiants and Asymptotic Giant Branch Stars: Comparisons with Isochrones}

	The number of objects in the CFHTIR fields increases significantly when $K > 16$, 
as expected given the peak RSG brightness measured by Rozanski \& Rowan-Robinson (1994). 
Thus, the majority of objects with $K > 16$ in Figures 2 and 3 are almost certainly RSGs 
and -- at fainter magnitudes -- bright AGB stars. The red colors of the majority of 
objects with $K > 16$ is consistent with them being highly evolved massive stars. 

	The $(M_K, (J-K)_0)$ CMDs of the East and North fields are compared with 
Z = 0.019 isochrones from Girardi et al. (2002) in Figure 4. A total line of 
sight reddening $E(B-V) = 0.3$ has been assumed (Perelmuter \& Racine 1995), with a 
distance modulus of 27.75 (Freedman et al. 2001). 
The peak of the observed RSG sequence comes within $0.1 - 0.2$ 
magnitudes of matching the peak of the log(t) = 7.0 models, and the brightest RSGs have 
$J-K$ colors that agree with the log(t) = 7.0 isochrones to within $\sim 0.1$ magnitudes. 
It is also clear from Figure 4 that when M$_K < -10$ the RSG plume is populated with 
stars that have ages that do not exceed $\sim 25$ Myr. Thus, stars brighter 
than M$_K = -10$ can be used to probe the star-forming history of the galaxy during 
the past few tens of millions of years. 

	The color distribution near M$_K = -10$ widens in both 
fields, and RSGs and AGB stars both occur when M$_K > -10$. The 
isochrones predict that these two sequences have different colors, such that the bluer 
stars in the CMDs are RSGs while the redder objects are evolving on the AGB. 
Hints of separate RSG and AGB sequences can be seen in the CMDs, although the 
errors in the photometry are such that it becomes difficult to distinguish between 
RSGs and AGB stars when M$_K > -9$.

	Older AGB stars dominate the faint end of the CMDs, 
forming a richly populated clump in the CMDs when $M_K > -9$. The main concentration 
of stars with M$_K \sim -8.5$ has an age log(t) $\sim 9.0$. This age is 
significant, as it matches the epoch of peak star formation in M82 (Parmentier 
et al. 2003), which is attributed to the first interaction between M82 
and M81. Thus, the CFHTIR data sample stars that span the time from within 
10 Myr of the present day to the onset of interactions between M81 and its companions.

\section{THE SPATIAL DISTRIBUTION AND SPECIFIC FREQUENCY OF RED SUPERGIANTS AND AGB STARS}

\subsection{The Spatial Distribution of RSGs}

	The comparisons with the isochrones in Figure 4 indicate that 
the majority of objects with M$_K$ between --10 and -- 12 are RSGs. 
The spatial distribution of objects with $H-K > 0.2$ 
in this brightness interval is investigated in Figure 5. The blue color cut-off was 
applied to avoid contamination from globular clusters. 
The panels in Figure 5 are positioned according to the actual 
location of each field on the sky, although the fields are shown 50\% 
larger than there actual size to facilitate the identification of individual stars.

	Broad trends can be seen in the spatial distribution of bright RSGs 
in Figure 5. In most fields there is a tendency for the number of RSGs to decrease 
with increasing distance from the center of M81, reflecting the overall distribution 
of stars in the disk. There is also a tendency for the RSGs to clump together, 
and a simple visual inspection suggests that this occurs on $\sim$ kpc spatial scales. 
This is not greatly different from the expected spatial distribution of stars 
with ages of $\sim 25$ Myr as they diffuse away from their place of birth (\S 5.1). 
The RSGs in the West field also appear to lie along a strip that runs diagonally across 
the field, which shows the spiral structure in this part of M81.

	The evidence for the clustering of RSGs notwithstanding, the brightest RSGs 
appear to be distributed throughout the fields, indicating that star formation during the 
past 25 Myr has not been restricted to a handful of locations; the CFHTIR 
fields are not dominated by single star-forming 
regions, but sample large-scale regions of star formation. The dispersed nature 
of the bright RSG population, coupled with the observation that the number 
of RSGs in each field is roughly similar, motivates us to make a field-to-field 
comparison of the density of RSGs. This is done in \S 4.2, where the number of 
stars per brightness interval per unit integrated brightness, which we refer to 
as the specific frequency, is computed.

\subsection{Field-to-field Comparisons of the Specific Frequency of RSGs}

	For this study, the specific frequency is defined as the number of objects 
per brightness interval in a system with a total integrated brightness M$_K = -16$. The 
2MASS Nearby Galaxy (Jarrett et al. 2003) $K-$band image of M81, which was smoothed with a 
$19 \times 19$ arcsec median boxcar filter to suppress individual sources, was used to 
compute the total $K-$band magnitude in each field, and the 
results are listed in the last column of Table 1. The integrated 
brightnesses of the four fields agree to within a few tenths of a magnitude, and this 
is a consequence of the prejudiced `by eye' procedure for selecting these 
fields, which identified regions of similar density near the edge of the M81 disk.

	The specific frequencies of the brightest stars in the four fields are compared in 
Figure 6, which shows the M$_K$ LF of each field, as constructed from the $(K, 
H-K)$ CMDs. Objects with $(H-K)_0 < 0.15$ have not been counted 
to avoid contamination from globular clusters (\S 3.2). The LFs 
have been corrected for incompleteness using the results from the artificial 
star experiments, and the error bars show the combined uncertainties in the raw 
counts and the completeness corrections.

	In \S 3.3 it was found that objects with 
M$_K < -10$ are expected to be RSGs, while objects with M$_K > -10$ may be either
RSGs or AGB stars; these magnitude regimes are marked in Figure 6. The LFs follow a rough 
power-law from M$_K = -11.5$ to fainter magnitudes, although there is the hint of a 
discontinuity in the LF near $M_K = -9$, which corresponds to the clump of AGB 
stars that are seen at the faint end of the CMDs. 

	The specific frequency of RSGs is highest in the North field, 
where the density of objects with M$_K$ between --10 and --11.5 falls 
$\sim 0.3$ dex above the average of all four fields. While the error bars in the 
individual bins of the North field LF are such that the number counts do not differ 
significantly from that of the mean curve, the $0.3$ dex difference is systematic over 
the four bins that span the upper RSG sequence, strongly suggesting that it 
is not a consequence of random errors. The comparison 
in Figure 6 thus suggests that the SFR during the 
past $\sim 25$ Myr was higher in the North field than in the other fields.

	The fractional contribution that AGB stars make to the star counts is 
expected to increase from M$_K = -10$ to --9, and the comparisons with the isochrones 
in Figure 4 indicate that the majority of objects in the last bin of the LFs, which have 
colors that are redder than those expected for RSGs, are evolving on the AGB. The rate 
at which AGB-tip brightness changes with age slows towards fainter 
magnitudes, and so the stars in the M$_K = -9$ LF bin span a wider range 
of ages than those in -- say -- the M$_K = -9.5$ bin. Because they contain stars covering a 
broader range of ages, the number counts in the faint bins probe a SFR that 
has been averaged over longer time scales than the LF bins that contain RSGs; thus, the 
faint end of the LFs in Figure 4 are less sensitive to temporal variations 
in the SFR, which one might expect to average out with time. 
In fact, there is a much smaller field-to-field dispersion in the specific 
frequencies of stars with M$_K = -9$ than at brighter magnitudes, as might be expected if 
the SFR is smoothed over long time intervals.

	The specific frequencies of stars in the CFHTIR data can be checked by 
making comparisons with the specific frequency of stars in 
the AOB data, which are less prone to crowding. There are 17 stars with M$_K$ between 
--9.25 and --8.75 in the AOB field, which covers an area of 1150 arcsec$^2$ 
in the West field. The $K-$band surface brightness in this portion 
of M81 was measured from the $19 \times 19$ arcsec$^2$ smoothed 2MASS $K-$band image of 
M81, and the specfic frequency of stars with M$_K$ between --9.25 and 
--8.75 is $11 \pm 3$. For comparison, the specific 
frequency measured in the entire CFHTIR West field in this same 
brightness interval is $\sim 6$, whereas the mean specific frequency in all four 
fields is $\sim 10$. Thus, the specific frequency of AGB stars measured from 
the AOB data agrees with that measured from the CFHTIR data within the $2-\sigma$ 
significance level.

\subsection{The Specific Frequency of RSGs Within Fields}

	The disks of spiral galaxies appear to have age gradients, in the 
sense that the luminosity-weighted age becomes younger with increasing radius 
(e.g. Bell \& de Jong 2000). Despite this trend, Davidge (2006) found that very young stars 
In the outer disk of NGC 247 eventually disappear at very large radii, and that beyond 
this point the ages of the youngest stars in a given distance interval increase with 
galactocentric distance. This could suggest that the SFR in the outer disk was 
higher in the past than at the present, or that stars that formed in the inner disk 
can be moved to larger radii on timescales of a few tens of Myr. To see if a similar 
trend occurs in M81, the East and West fields, where the greatest range of distances are 
sampled given the inclination of the M81 disk to the line of sight, were divided in half; 
in the subsequent discussion the half that is closest to the galaxy center is called 
the `inner' portion of the field, while the half that is furthest from the galaxy center 
is called the `outer' portion. In each field the inner portion samples galactocentric 
radii between 4 and 7 kpc, while the outer portion samples radii between 7 and 10 kpc.

	The specific frequencies of stars in the inner and outer portions of the 
East and West fields are compared in Figure 7. The comparisons in Figure 7 reinforce the 
results from Figure 5, which indicate that RSGs are present in the outer portions of both 
fields; therefore, any edge to the young M81 disk occurs outside the range of radii covered 
in these fields. Moreover, when the summed data from the inner and outer portions of 
both fields are considered it can be seen that -- with the exception of RSGs with 
M$_K = -10.5$ -- the specific frequency of RSGs does not change with radius. 
The clear excess number of stars at M$_K = -10.5$ in the inner portion of both 
fields suggests that the SFR per unit disk mass was higher $\sim 20$ Myr in the 
past when R$_{GC} < 7$ kpc than at larger radii.

\section{DISCUSSION \& SUMMARY}

	Near-infrared images that were recorded with the CFHT have been used to 
investigate the properties of RSGs and AGB stars in the disk of M81. Four fields that 
each subtend a projected area of $3.9 \times 3.9$ kpc$^2$ were observed. 
When considered together the four fields sample a 
total integrated brightness $K = 6.0$, which amounts to 13\% of 
the light from M81. Thus, there is a reasonable expectation that these data sample 
areas that are representative of the M81 disk at intermediate galacocentric distances.

	The densities of RSGs and AGB stars between and within the fields 
have been investigated by computing the number of objects per magnitude interval, 
normalized to a system with M$_K = -16$, which is referred to as the specific frequency. 
RSGs with ages $< 25$ Myr are seen throughout the four fields. While the 
specific frequency of bright RSGS is similar in three of the four fields, the 
specific frequency of these objects is highest in the northern part of the 
galaxy. These results suggest that widespread star formation occured throughout much 
of the disk of M81 during the past $\sim 25$ Myr, with pockets of localized intense 
activity. This is very different from what is seen today, where star formation is 
clearly concentrated at intermediate distances along the spiral arms (Gordon et al. 2004).

\subsection{Comparing the Locations of Star Formation in the Present and 25 Myr in the Past}

	Studies of resolved stars can be used to trace the star-forming histories 
of galaxies, and identify areas of on-going and past star 
formation; with information of this nature then the movement of star-forming 
activity throughout a galaxy can be charted. In the case of M81, widespread elevated 
levels of star formation are expected to have occured at the time when the interactions 
with M82 and NGC 3077 were most intense. Given that this likely happened at least a few 
hundred million years in the past and that the timescale for galaxy-wide star formation 
is $\sim 0.1$ Gyr (e.g. Efremov \& Elmegreen 1998), then (1) the current SFR is almost 
certainly lower than the mean rate over the past few hundred million years, and (2) the 
areas of current active star formation may not be the same as during previous epochs. 

	Three of the CFHTIR fields sample areas of active 
star formation at the present day. The South field samples the dominant concentration of 
bright mid-IR flux found by Gordon et al. (2004) in the southern half of 
M81, and Figure 2 of Gordon et al. (2004) indicates that this is also an area of 
bright UV and H$\alpha$ emission. The North field samples the portion of the 
northern spiral arm that is a strong source of mid-IR emission, as well as 
strong UV and H$\alpha$ emission. The West field samples 
the next brightest region of mid-IR emission in M81, which is also an area of prominent 
UV and H$\alpha$ emission. For comparison, only a few weak sources of mid-IR emission are 
located in the East field; while there is strong UV 1500\AA\ emission in the East field, 
there is only weak emission at longer UV wavelengths and in H$\alpha$. The 
H$\alpha$ image shown in Figure 3 of Lin et al. (2003) confirms the regions of 
star-forming activity indicated by the mid-IR images.

	The mid-IR measurements made by Gordon et al. (2004) indicate that star formation 
at the present day tends to be concentrated at intermediate galactocentric 
radii in the northern and southern portions of M81.
Boissier et al. (2003) investigated the radial distribution 
of matter in M81, and found that while M81 has a relatively high 
SFR, the radially-averaged gas density is subcritical 
for triggering star formation throughout the disk. The projected density of molecular 
material in M81 is also relatively low (Brouillet et al. 1998). The radially-averaged gas 
density in M81 comes closest to the critical threshold for star formation at a 
projected radius of $\sim 9$ kpc, and so this is where star-forming activity might be 
expected to peak. However, star-forming activity tends to occur at roughly one half 
this radius, and the areas of current star formation are likely regions 
where the local gas density breaches the critical threshold due to, for example, the 
passage of spiral density waves. 

	That RSGs with M$_K < -10$ are seen throughout each CFHTIR field 
suggests that star formation in the disk of M81 during the past $\sim 25$ Myr 
occured over larger areas than at present. The comparisons in Figure 6 indicate that 
the specific frequency of the brightest RSGs in the South, East, and West fields is 
very similar, indicating that the star-forming histories of these fields during 
the past $\sim 10 - 25$ Myr have also been similar, even though the present-day 
SFR in the East field appears to be much lower than in the South and West fields. 

	A smooth distribution of stars can be produced if they are 
scattered from their places of birth via interactions 
with massive gas clouds. However, the similar specific frequencies of 
bright RSGs in the East, South, and West fields is almost certainly not due to such a 
process. Consider the Galactic disk as an example. The velocity 
dispersion of solar neighborhood Cepheids, which have 
ages comparable to that of RSGs, is $\sim 10$ km sec$^{-1}$ (Wielen 1974). 
Objects with this velocity will travel only $\sim 0.3$ kpc during 25 Myr. While the 
applicability of the dynamics of stars in the Galactic disk to 
M81 may seem dubious given the possible differences between the molecular 
cloud contents of the Milky-Way and M81 (Brouillet et al. 1998; Taylor \& 
Wilson 1998), it is significant that the clustering size of bright RSGs in 
Figure 5 is not greatly different from the spatial scattering distance predicted from 
Milky-Way Cepheids.

	Unless RSGs in the disk of M81 are dispersed much more quickly than stars 
in the Galactic disk then the uniform specific frequency 
of RSGs in three of the four CFHTIR fields suggests that the 
local SFR in the disk of M81 at a projected galactocentric radius of 5 kpc, normalized 
per unit projected mass density as probed by integrated $K-$band light, has been 
constant when averaged over $\sim 25$ Myr time spans. It should be emphasized that 
this conclusion likely holds only for a strip with a width of a few 
kpc at R$_{GC} \sim 5 - 7$ kpc. Indeed, the comparisons in Figure 6 hint that 
there was more active star formation when R$_{GC} < 7$ kpc, which presumably causes 
the high specific frequency of RSGs with M$_K \sim -10.5$ when R$_{GC} < 7$ kpc. The 
absence of HII regions when R$_{GC} < 3$ kpc (e.g. Lin et al. 2003) also indicates 
that star formation at the current day does not occur in the inner disk of M81. 

	Despite the tendency for more distributed star formation in M81 10 -- 25 Myr 
in the past, there are areas of star formation that have stayed active for at least this 
period of time. Indeed, the specific frequency of RSGs in the North field, where there 
is vigorous star formation at the current day, is two times that in the other fields. 
This is thus an area of prolonged star-forming activity.

\subsection{Estimating the SFR in M81 25 Myr in the Past}

	The data discussed in this paper suggest that M81 contains a rich population of 
very luminous RSGs. The specific frequency measurements indicate 
that RSGs brighter than M$_K = -10$ account for 1.5\% of the $K-$band light. If 
M81 is assumed to be a typical Sb galaxy, with 16\% of its light coming from the 
bulge (Perez-Gonzalez et al. 2001), then there should be $\sim 3300$ RSGs 
with M$_K < -10$ in the disk of M81. 

	Using data listed in Table 4 of Bell \& Kennicutt (2001), and the 
calibrations given in their Equations 2 and 3, the present-day SFR of M81 is 0.24 
M$_{\odot}$ year$^{-1}$ based on the UV flux, and 0.36 M$_{\odot}$ year$^{-1}$ based 
on the H$\alpha$ flux. How does this compare with the SFR $\sim 10 - 25$ Myr in the 
past? This question can be answered by estimating the mean SFR over the 
interval 10 -- 25 Myr from the specific frequency of RSGs. If there are 
3300 RSGs brighter than M$_K = -10$ in M81, and if these stars have progenitor 
masses between 10 and 20 M$_{\odot}$, then the RSG progenitors had a total mass $\sim 5 
\times 10^4$ M$_{\odot}$. An integration of the Kroupa et al. (1993) mass function 
with limiting masses of 0.08 and 100 M$_{\odot}$ 
indicates that stars in the 10 -- 20 M$_{\odot}$ range account for 
$\sim 3\%$ of the total mass formed. The SFR for the disk of M81 between 10 and 25 
Myr in the past is then $\sim 0.1$ M$_{\odot}$ year$^{-1}$. The estimated SFR does not 
change markedly if a Salpeter (1955) mass function is assumed. We conclude that while star 
formation may have been distributed over larger areas of M81 than today, the mean SFR in 
M81 was not markedly higher over the past 25 Myr.

\clearpage

\begin{table*}
\begin{center}
\begin{tabular}{lccc}
\tableline\tableline
Field & RA & Dec & K \\
\tableline
M81 West & 09:54:44 & 69:05:45 & 7.32 \\
M81 East & 09:56:24 & 69:03:54 & 7.65 \\
M81 North & 09:55:24 & 69:10:11 & 7.84 \\
M81 South & 09:56:04 & 68:58:16 & 7.39 \\
\tableline
\end{tabular}
\end{center}
\caption{Locations of the M81 Fields}
\end{table*}

\clearpage

\begin{table*}
\begin{center}
\begin{tabular}{lrcccc}
\tableline\tableline
ID & [Fe/H] & $K$ & $H-K$ & $J-K$ & $V-K$ \\
\tableline
Id50401 & --0.04 & 17.10 & 0.18 & 0.76 & 2.83 \\
Id50415 & --1.90 & 17.00 & 0.19 & 0.74 & 2.24 \\
Id50552 & 0.98 & 17.18 & 0.18 & 1.08 & 2.34 \\
Id50785 & --0.72 & 16.24 & 0.28 & 1.06 & 2.84 \\
\tableline
\end{tabular}
\end{center}
\caption{Photometry of Spectroscopically Confirmed Globular Clusters}
\end{table*}

\clearpage

\begin{figure}
\figurenum{1}
\epsscale{0.75}
\plotone{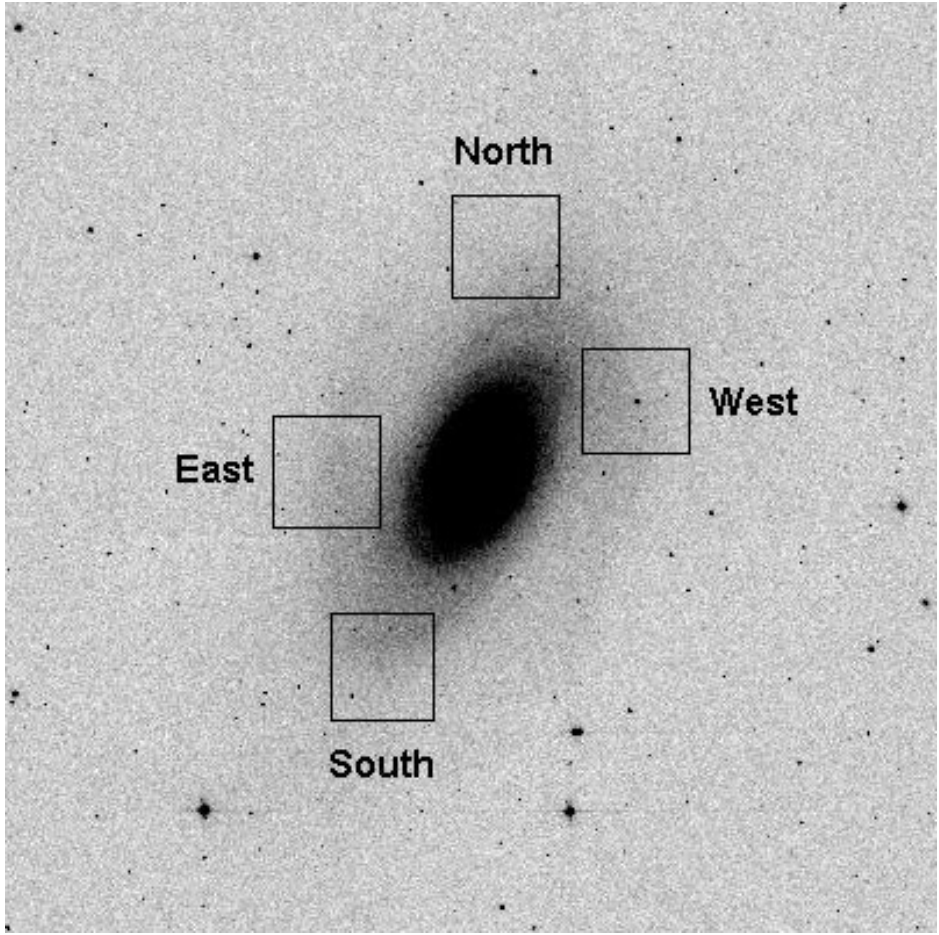}
\caption
{The locations of the CFHTIR fields are indicated on this $27 \times 27$ arcmin$^2$
$K-$band image from the 2MASS Large Galaxy Atlas (Jarrett et al. 2003). The bright star 
at the center of the West field is GSC 04383--00308, which defines the center of the 
CFHT AOB field.}
\end{figure}

\begin{figure}
\figurenum{2}
\epsscale{0.75}
\plotone{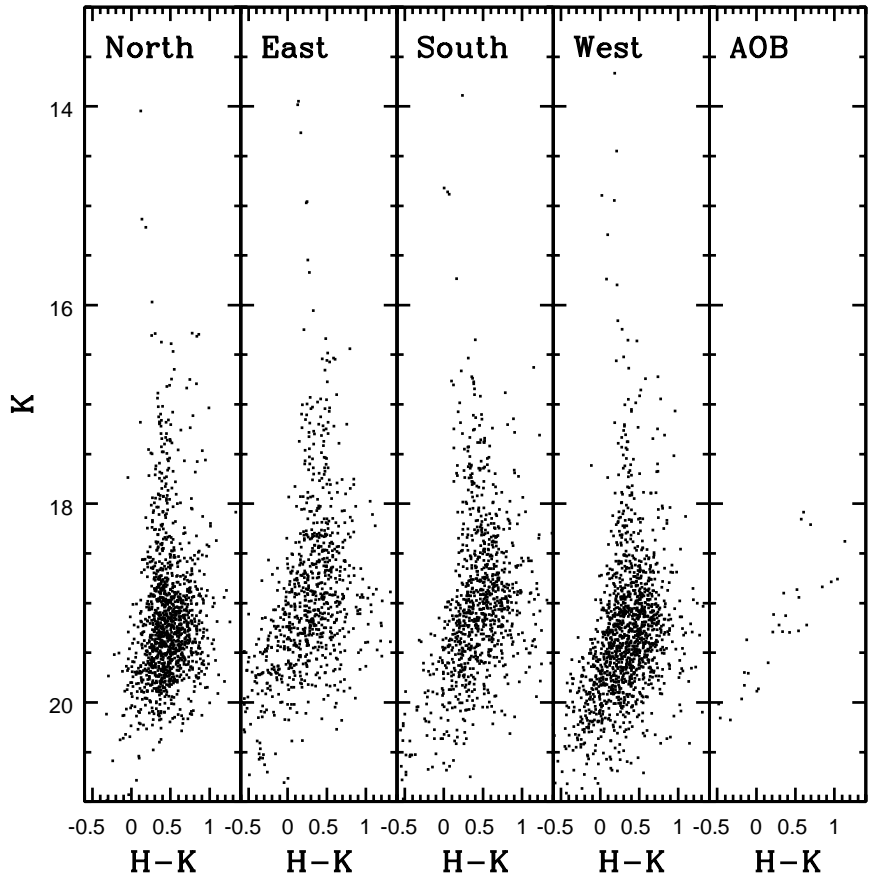}
\caption
{The $(K, H-K)$ CMDs of the M81 fields. Individual stars in M81 
have $K > 16$; RSGs populate the red plume of objects with $K$ between 16 and 18, 
while AGB stars dominate when $K > 18.5$. Some of the sources with bluer colors are 
known globular clusters (\S 3.2).}
\end{figure}

\begin{figure}
\figurenum{3}
\epsscale{0.75}
\plotone{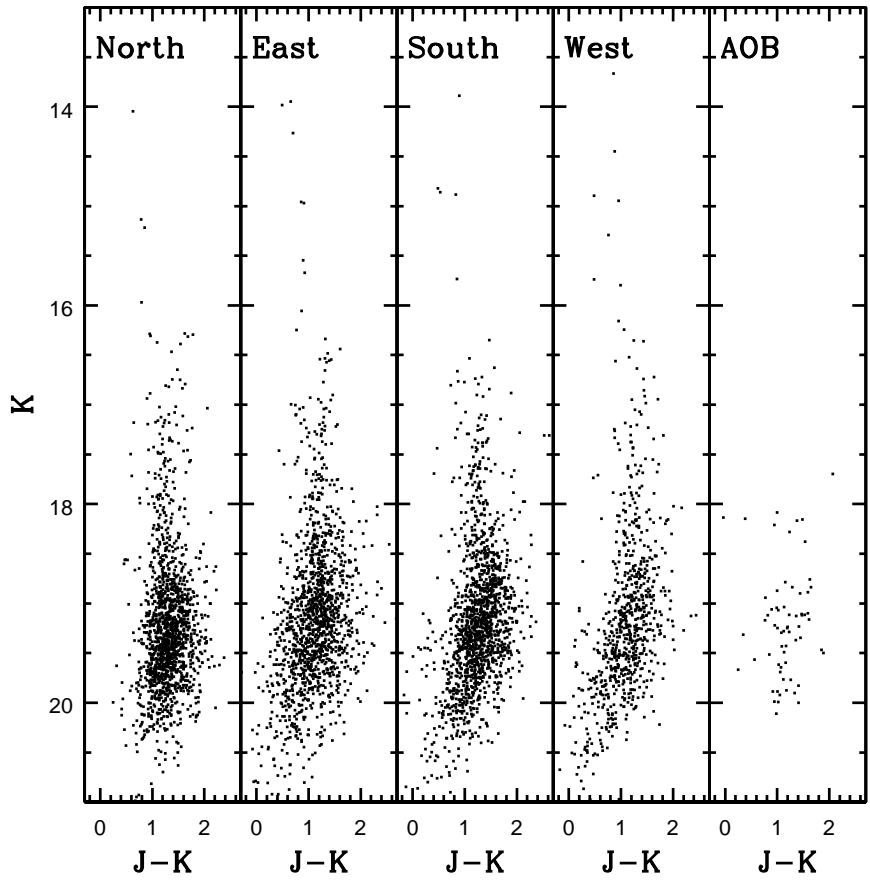}
\caption
{The same as Figure 2, but showing the $(K, J-K)$ CMDs.}
\end{figure}

\begin{figure}
\figurenum{4}
\epsscale{0.75}
\plotone{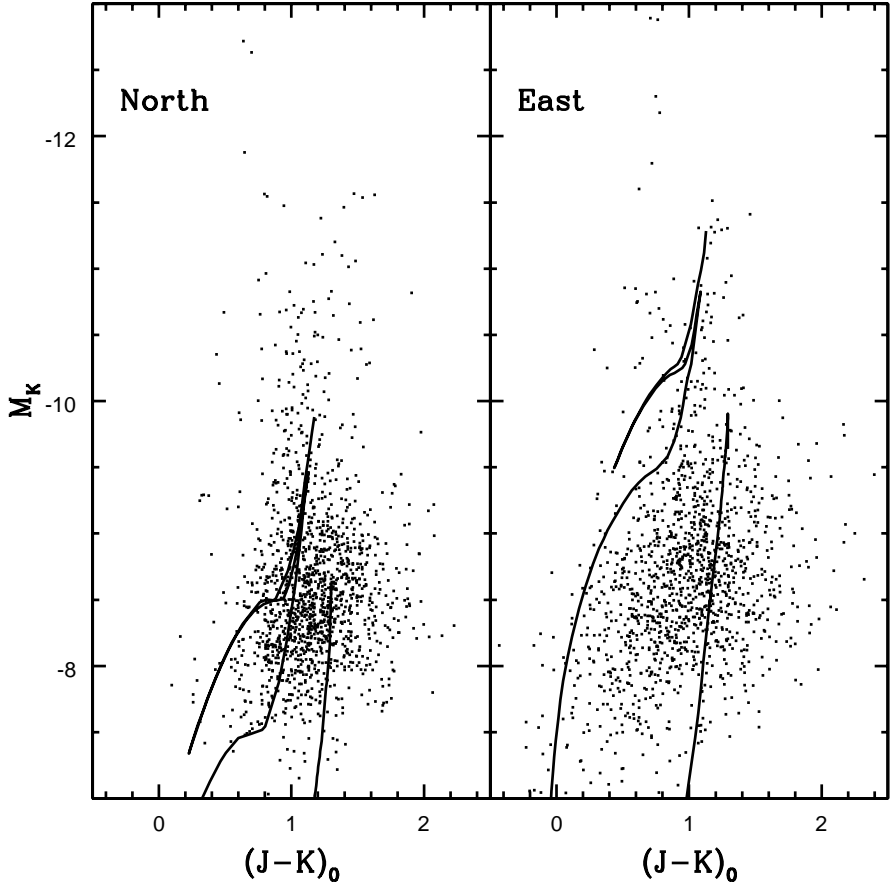}
\caption
{The $(M_K, (J-K)_0)$ CMDs of the North and East fields are compared with 
Z=0.019 models from Girardi et al. (2002). The sequences 
in the left hand panel have log(t$_{yr}) = 7.4$, and 9.0, while those 
in the right hand panel have have log(t$_{yr}) = 7.0$ and 8.1. A distance modulus of 
27.75 and $E(B-V) = 0.3$ have been assumed. Note that the magnitudes and colors of 
the brightest red stars are roughly matched by the log(t) = 7.0 isochrone. 
The reddest stars with M$_K < -10$ can be matched with the log(t) = 8.1 model. This 
particular age marks the onset of the AGB, and the isochrone comes within a few tenths 
of a magnitude of matching the peak brightness of the main body of red stars in the 
East field. The majority of the stars that form the clump near the faint end 
of the CMDs are evolving on the AGB and have log(t) $\sim 9$.}
\end{figure}

\begin{figure}
\figurenum{5}
\epsscale{0.75}
\plotone{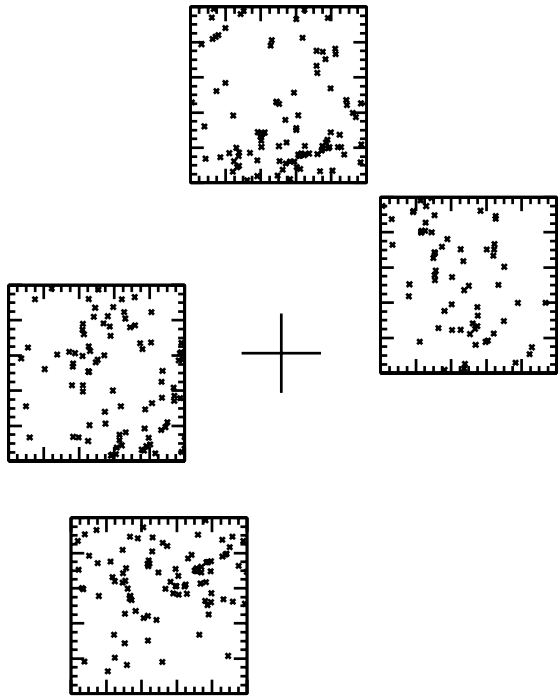}
\caption
{The spatial distribution of stars with M$_K$ between --10 and --12, which the 
isochrones in Figure 4 indicates is the brightness interval that is dominated 
by RSGs. Sources with $H-K < 0.2$ were not counted, as this color range is dominated by 
globular clusters (\S 3.2). The location of the boxes in this figure tracks the 
positions of the CFHTIR fields on the sky, although the 
fields are shown 50\% larger than their true size to prevent the crowding of data points. 
The center of M81 is indicated with a cross. Note that while there is a tendency 
for stars to clump on roughly kpc scales, bright RSGs are seen throughout 
much of the CFHTIR fields.}
\end{figure}

\begin{figure}
\figurenum{6}
\epsscale{0.75}
\plotone{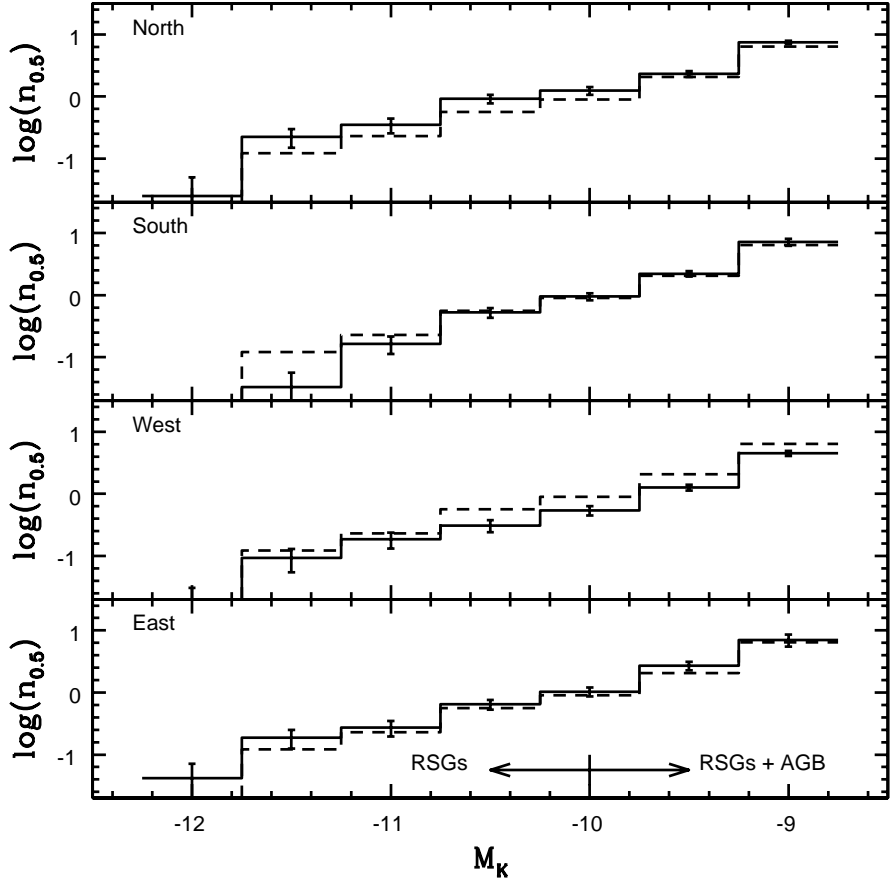}
\caption
{The $K$ LFs of stars in the four M81 fields (solid lines) are compared with the mean LF 
of all four fields (dashed lines). The LFs were constructed from the $(K, H-K)$ CMDs, 
and have been normalized to the counts expected if each field samples an integrated 
brightness M$_K = -16$. $n_{0.5}$ is the number of stars per 0.5 magnitude 
interval in a system with M$_K = -16$. Sources with $H-K < 0.2$ were not 
counted, as many of these are likely globular clusters (\S 3.2). The error bars show the 
combined uncertainties in the raw number counts and in the completeness corrections. 
The magnitude intervals occupied by RSGs and a mixture of RSGs and AGB stars, based on the 
isochrones in Figure 4, are also indicated. Note that the specific frequency of 
RSGs in the North field is systematically higher than average. 
This is suggestive of differences in the local SFR over kpc or larger spatial scales 
in the M81 disk during the past $\sim 25$ Myr. The specific frequencies of 
stars at the faint end, where AGB stars are expected to dominate, tend to 
show less field-to-field dispersion than at the bright end.}
\end{figure}

\begin{figure}
\figurenum{7}
\epsscale{0.75}
\plotone{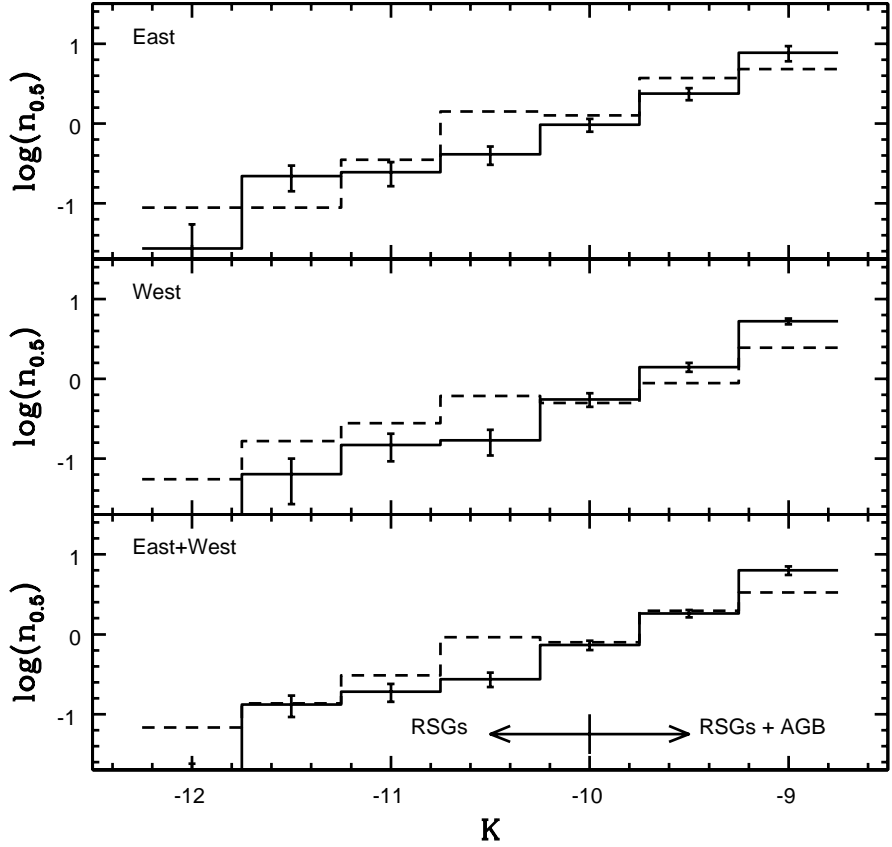}
\caption
{The $K$ LFs of stars in the outer portions of the East and West fields (solid lines) 
are compared with the LFs of the inner portions (dashed lines). The LFs produced by 
summing those of the inner and outer portions of both fields are compared 
in the lower panel. The LFs in this figure were constructed using the same 
procedures employed to generate the LFs in Figure 5, and the error bars show 
the combined uncertainties in the raw number counts and in the completeness corrections. 
The intervals occupied by RSGs and a mixture of RSGs and AGB stars, based on the 
comparisons with isochrones in Figure 4, are also indicated. With the exception 
of the M$_K = -10.5$ bin, the specific frequency of RSGs does not change significantly 
with radius throughout the East and West fields.} 
\end{figure}

\end{document}